\documentclass[prb,twocolumn,aps]{revtex4-1}
\usepackage{graphicx}
\usepackage{color}
\usepackage{amsmath}

\begin{document}

\def\K{{\bf{K}}}
\def\Q{{\bf{Q}}}
\def\X{{\bf{X}}}
\def\Gbar{\bar{G}}
\def\tk{\tilde{\bf{k}}}
\def\k{{\bf{k}}}
\def\q{{\bf{q}}}
\def\x{{\bf{x}}}
\def\y{{\bf{y}}}

\title{Quantum Criticality Due to Incipient Phase Separation in the Two-dimensional Hubbard Model}

\author{E. Khatami,$^{1,2}$ K. Mikelsons,$^{1,2}$ D. Galanakis,$^{1}$ 
A. Macridin,$^{3}$ J. Moreno,$^{1}$ R. T. Scalettar,$^{4}$ and M. Jarrell$^{1}$}
\affiliation{$^{1}$Department of Physics and Astronomy, Louisiana State University, Baton Rouge, Louisiana, 70803, USA\\
$^{2}$Department of Physics, University of Cincinnati, Cincinnati, Ohio, 45221, USA\\
$^{3}$Fermilab, P. O. Box 500, Batavia, Illinois, 60510, USA\\
$^{4}$Department of Physics, University of California, Davis, California 95616, USA}

\date{\today}

\begin{abstract}

We investigate the two-dimensional Hubbard model with next-nearest-neighbor hopping, $t^\prime$,
using the dynamical cluster approximation. We confirm the existence of a first-order
phase-separation transition terminating at a second-order critical point at filling
$n_c(t^\prime)$ and temperature $T_{ps}(t^\prime)$. We find that as $t^\prime$
approaches zero, $T_{ps}(t^\prime)$ vanishes and $n_c(t^\prime)$ approaches the
filling associated with the quantum critical point separating the Fermi liquid from
the pseudogap phase. We propose that the quantum critical point under the superconducting
dome is the  zero-temperature limit of the line of second-order critical points.
\end{abstract}

\maketitle

%==========BODY OF PAPER =========================================
\section*{Introduction}

Strongly correlated electronic materials, which include high-temperature
superconductors, heavy fermions, and magnetic compounds, are
characterized by competing phases and complicated phase diagrams. These
competing phases can lead to quantum criticality when one of the
transition temperatures  is driven to absolute zero as a function of a
non-thermal control parameter such as pressure or
doping~\cite{p_coleman_05,s_sachdev_08}.  While the physics of a
conventional phase transition is driven by thermal fluctuations, near a
quantum critical point (QCP), quantum fluctuations affect the properties
of a material up to surprisingly high temperatures \cite{a_kopp_05}.  In
particular, transport measurements of hole-doped cuprates suggest the
presence of a QCP~\cite{dirk,r_daou_09} lying beneath the superconducting (SC)
dome.~\cite{f_Balakirev_09,varma_physrep} Although it is believed that this QCP
dominates the phase diagram, its nature is still unknown\cite{broun08}
with competing scenarios emphasizing the role of bosonic or fermionic
fluctuations.\cite{dirk}

\begin{figure}[t]
\begin{center}
\includegraphics*[width=0.37\textwidth]{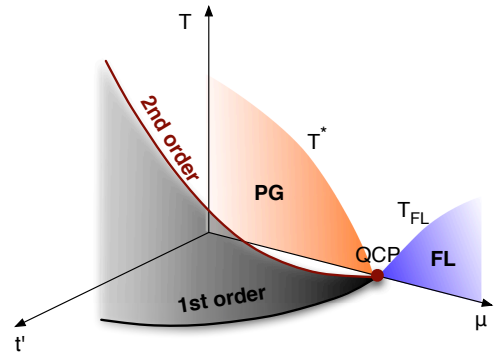}
\caption{(Color online) Schematic phase diagram of the 2D Hubbard
model in the temperature ($T$), chemical potential ($\mu$), and 
next-near-neighbor hopping  $(t^\prime$) space.
For $t^\prime>0$ the first-order phase-separation transition terminates 
at a second-order critical point at doping $n_c$ and temperature
$T_{ps}$. The line of second-order critical points $(T_{ps},n_c)$, 
approaches the QCP on the $t^\prime=0$ plane. 
This is the critical point separating the pseudogap 
(PG) from the Fermi-liquid (FL) region.}
\label{fig:scematic_phase_diagram}
\end{center}
\end{figure}

In this work, we provide evidence for the nature of the QCP in the
Hubbard model of the cuprates by a systematic study of its phase
diagram. Our results suggest that the QCP is {\em{not}} due to
order in the pseudogap (PG) region, but rather is the zero-temperature
limit of a line of second-order critical points associated with a first-order 
phase-separation transition (see
Fig.~\ref{fig:scematic_phase_diagram}).  The control parameter for this
transition is the next-near-neighbor hopping parameter, $t^\prime$. 

Although it is possible to have a QCP not associated with any obvious
order parameter, as in the case of a localization transition, in most
QCPs a continuous order parameter vanishes at $T=0$ for a particular
value of the controlling energy scale~\cite{vojta_03}.  Less common is a
QCP associated with a first-order transition, but this is possible when
the first-order transition terminates at a second-order critical point
which is driven to zero by tuning the relevant parameter. For
example, in Sr$_3$Ru$_2$O$_7$~\cite{grigera}, as a function of the field
angle, a first-order meta-magnetic transition is driven to $T=0$
yielding quantum critical phenomena.

The latter case is consistent with the scenario of
Fig.~\ref{fig:scematic_phase_diagram}, where the role of the field angle
is played by $t^\prime$. For positive $t^\prime$ and below a certain
temperature, the system undergoes a first-order phase-separation
transition. In this region, two solutions with different densities
coexist for a given chemical potential, $\mu$~\cite{alex_PS,m_Aichhorn_07}.
More recently, this first-order metal-insulator transition was studied by 
Gull {\em et al.}~\cite{gull} and used to map out the phase diagram of this 
model in the space of interaction strength and $t'$.
Since these two phases have the same symmetry, this transition terminated
in a second-order critical point at temperature $T_{ps}$ and critical
filling $n_c$ where the charge susceptibility diverges.  By increasing
$t^\prime$, $T_{ps}$ increases and the critical point becomes numerically
accessible. For low $t^\prime$, it is no longer accessible, but its
presence is evidenced by a peak in the charge susceptibility.

Our starting point is the two-dimensional (2D) Hubbard Hamiltonian,
\begin{equation}
H=\sum_{\k\sigma}\epsilon_{\k}^{0}c_{{\k}\sigma}^{\dagger}
c_{{\k}\sigma}^{\phantom{\dagger}}+U\sum_{i}n_{{i}\uparrow}n_{{i}\downarrow},
\label{eq:hubbard}
\end{equation}
where $\epsilon_{\k}^{0}=-2t\left(\cos k_x+\cos k_y\right)
-4t^\prime\left(\cos k_x \cos k_y -1 \right)$
is the tight-binding dispersion as a function of the hopping $t$ between
nearest neighbors and $t^\prime$ between next-nearest neighbors, 
$c_{{\k}\sigma}^{\dagger}(c_{{\k}\sigma})$ is the 
creation (annihilation) operator for electrons of wave vector ${\k}$ and 
spin $\sigma$, $n_{i\sigma}=c_{{i}\sigma}^{\dagger}c_{{i}\sigma}$ and $U$ 
is the on-site Coulomb repulsion.

\section*{Methodology}
We solve the Hubbard model within the dynamical cluster approximation 
(DCA)~\cite{hettler:dca} on a $N_c$-site cluster.  The DCA is a cluster 
mean-field theory which maps the original lattice model onto a periodic
cluster of size $N_c=L_c^2$ embedded in a self-consistent host. Spatial 
correlations up to a range $L_c$ are treated explicitly, while those at 
longer length scales are described at the mean-field level.   However the 
correlations in time, essential for quantum criticality,  are treated
explicitly for all cluster sizes.  To solve the cluster problem, we use 
weak-coupling expansion continuous time quantum Monte Carlo (QMC)
method~\cite{rombouts99,rubtsov05}
 with highly optimized blocked and delayed updates~\cite{karlis}, a
determinant QMC method which scales linearly in the inverse
temperature~\cite{ehsan}, as well as Hirsch-Fye QMC~\cite{hirsch86,jarrelldca}.  
The fast determinant QMC was used to obtain a converged DCA solution, 
Hirsch-Fye QMC was used to calculate lattice susceptibilities, and 
continuous time QMC was used as a control
for systematic error.  The unit of energy is $t$ in the entire paper.  

To make contact with previous results, we perform simulations with $U=6$, but we 
find that the phase separation becomes more prevalent for $U=8$ for 
which we present most of our results.  We calculate the filling, $n$, versus 
$\mu$ and the compressibility, $dn/d\mu$, by taking the numerical derivative.  
We also calculate various susceptibilities including the charge $\chi_c(\Q=0,T)$,
spin and pairing susceptibilities by solving the lattice Bethe Salpeter equation using 
the renormalized DCA vertices \cite{jarrelldca}.  Note that as a
consequence of the fluctuation-dissipation theorem,
$\chi_c(\Q=0,T)$ is identical to $dn/d\mu$, but we keep both terms,
compressibility and susceptibility, to identify the method by which
they are calculated.

\section*{Results} 

\begin{figure}[t]
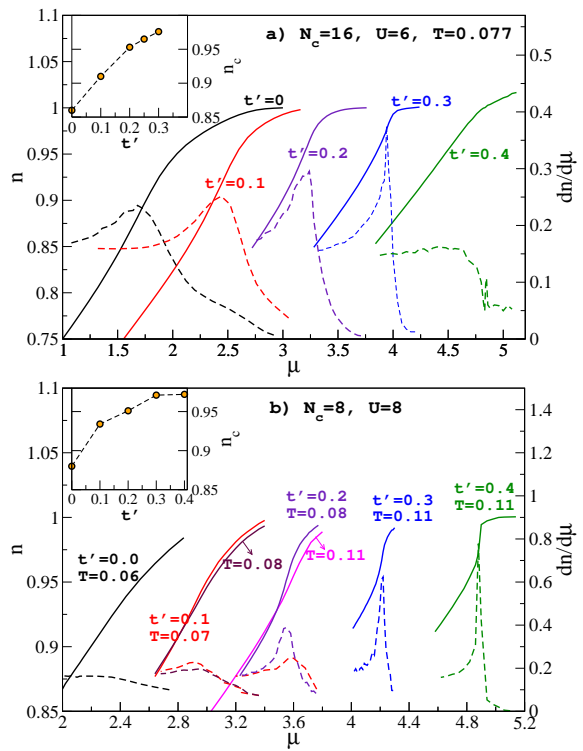

\begin{center}
\includegraphics*[width=0.42\textwidth]{fill_of_mu_Nc16B_U1.5_tp0.0-0.3.eps}
\includegraphics*[width=0.42\textwidth]{Nvsmu_Nc8A_U8.eps}
\caption{(Color online) Filling, $n$ (solid lines), and compressibility,
$dn/d\mu$ (dashed lines), plotted vs chemical potential, $\mu$, for 
various values of $t^\prime$ for (a) $U=6$, $N_c=16$, and
$T=0.077$, and (b) $U=8$, $N_c=8$  at different temperatures. The unit of energy is
$t$ in all figures. The critical filling, where the compressibility
peaks, is plotted in the corresponding inset.  In (a) when
$t^\prime\to0$ the peak in the charge susceptibility is located at the
QCP identified previously~\cite{raja}.}
\label{fig:nmut}
\end{center}
\end{figure}

In Fig.~\ref{fig:nmut}(a), we plot $n$ versus $\mu$ for $U=6$, 
$T =0.077$ and different values of $t^\prime$, ranging from $0.0$ to $0.4$, on a 16-site cluster. 
The filling increases monotonically with the chemical potential and shows a pronounced 
flat region, associated with the Mott gap, especially for
$t^\prime<0.4$. The most interesting feature of  $n(\mu)$ is an 
inflection  apparent at finite doping, which becomes more pronounced for 
larger values of $t^\prime$. The inflection in $n(\mu)$ translates into 
a peak in the compressibility (shown on the right axis).
The peak becomes sharper and moves closer to half filling as
$t^\prime$ increases. For $t^\prime>0.3$, the
peak disappears, as does the plateau in $n(\mu)$ near half filling, 
associated with the gap. The value of the critical filling at the peak, $n_c$, 
versus $t^\prime$ is plotted in the inset.
Note that for $t^\prime=0$, $n_c=0.86$ agrees with the filling $0.85$ of the QCP
identified previously for these parameters~\cite{raja,karlisQCP} and 
a critical filling separating two Fermi-liquid (FL) regions in a closely 
related $t-J$ model.~\cite{k_haule_07}
These results suggest that the QCP may be associated with charge fluctuations.

To explore this association, we study the behavior of the bulk 
charge susceptibility, $\chi_c(\Q=0,T)$ and its divergence as $t^\prime\to 0$. 
Unfortunately, for this cluster and parameters, the minus sign
problem~\cite{loh_sign_problem}
limits our ability to access temperatures low enough to see a divergence 
in the charge susceptibility.  The minus sign problem becomes worse when 
the cluster size, $U$ or $\left |t^\prime\right|$ increases or the temperature
decreases.  However, the cluster size, interaction $U$ and $t^\prime$ 
affect the phase diagram in different ways.  In previous 
studies~\cite{alex_PS}, we found that for $t^\prime=0.3$, clusters with $N_c=8,12$ 
and $16$ have roughly the same phase-separation transition temperature when 
$U=8$.  Here, using the same interaction strength, we find that the  
cluster size effects are stronger for smaller values of $t^\prime$. As the 
cluster size is decreased, the charge susceptibility is somewhat suppressed, 
and more significantly, the critical doping moves towards half filling.
With increasing $U$ in the range 
from $U=4$ to $8$, the peak in the charge susceptibility moves to lower 
fillings and higher temperatures. So, despite the worse minus sign 
problem associated with larger values of the interaction, the region of 
divergent charge fluctuations becomes larger and more accessible for $U=8$.   
For this reason, from this point on we will use a smaller cluster with $N_c=8$ and $U=8$,
in order to access the second order critical points and investigate
their relationship to superconductivity.

In Fig.~\ref{fig:nmut}(b), we plot $n$ and $dn/d\mu$ versus 
$\mu$ for several $t^\prime$ and temperatures. Similar to  
Fig.~\ref{fig:nmut}(a), we see a cusp emerge in the compressibility. 
As the temperature is lowered, the peak in the compressibility is enhanced
for $t^\prime=0.1$ and $0.2$. However, for a larger $t^\prime$, e.g., $t^\prime=0.3$, 
and for $T<0.1$, we find hysteresis 
between two stable solutions with different values of $n$ for the same value of 
$\mu$~\cite{alex_PS}. The presence of hysteresis indicates that the system 
has undergone a first-order phase-separation transition. 

\begin{figure}[t]
\begin{center}
\includegraphics*[width=0.42\textwidth]{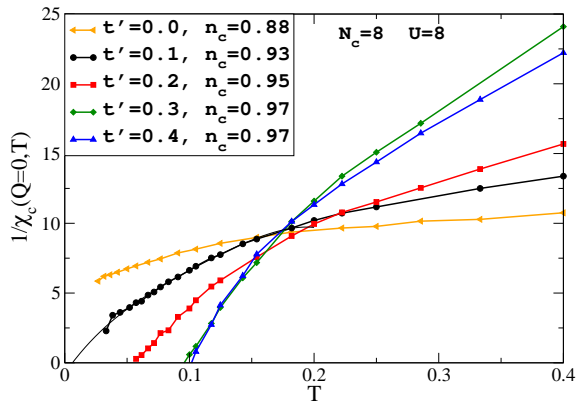}
\caption{(Color online) Inverse bulk charge susceptibility
vs temperature when $U=8$, $N_c=8$ for several values of 
$t^\prime$.  The values of the critical filling $n_c$ shown in the legend
correspond to the maximum of low temperature compressibility, or the filling 
where it first diverges.}
\label{fig:charge}
\end{center}
\end{figure}

We explore the line of second-order critical points of these first-order 
transitions as $t^\prime$ changes
using the charge susceptibility as shown in Fig.~\ref{fig:charge}.
Here, the inverse charge susceptibility at $n_c$ is plotted 
versus temperature for different values of $t^\prime$ when $U=8$ and 
$N_c=8$.  The critical filling identified in the legend is 
determined as the filling where the compressibility either diverges, or 
is peaked at the lowest accessible temperatures.
We find that the temperature of the second-order critical point increases
with increasing $t^\prime$ and that it moves towards half filling.
However, unlike the $U=6$ results shown in Fig.~\ref{fig:nmut}(a), the critical point appears to
avoid half filling even for $t^\prime=0.4$. The stronger Coulomb 
interaction $U=8$ also strengthens the Mott gap for this $t^\prime$ as can be seen in
the persistence of the flat region in $n(\mu)$ near $n=1$ for
$t^\prime=0.4$ [Fig.~\ref{fig:nmut}(b)].

\begin{figure}[t]
\begin{center}
\includegraphics*[width=0.47\textwidth]{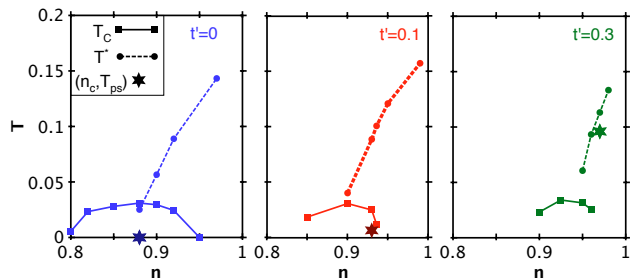}
\caption{(Color online) SC transition temperature $T_c$ (solid lines),
pseudogap temperature $T^*$ (dotted lines), and critical points (stars)
for the 2D Hubbard model with $U=8$, $N_c=8$, and $t^\prime=0$, $0.1$,
and $0.3$. For all $t^\prime$ the $T^*$ line terminates inside the SC dome.  
For $t^\prime=0$,  this termination point coincides with the
QCP of the phase-separation transition (star in the left panel).
For $t^\prime=0.1$ and $0.3$ the second-order critical point is at critical 
filling $n_c=0.93$ and $0.97$,
respectively, which is above the SC optimal filling.}
\label{fig:phase_Nc8_U8}
\end{center}
\end{figure}

The charge fluctuations associated with phase-separation influence the SC 
phase diagram.  This is shown in Fig.~\ref{fig:phase_Nc8_U8} for $N_c=8$, 
$U=8$, and $t^\prime=0.0$, $0.1$, and $0.3$. 
The pseudogap temperature, $T^*$, obtained as the temperature where 
the bulk spin susceptibility peaks (see Ref.~\onlinecite{raja}), is also plotted.
For $t^\prime=0$, $T^*$ vanishes at the QCP, 
which for this smaller cluster has moved to $n_c=0.88$.  Note that for $t^\prime=0$,
the SC dome is centered on the QCP, suggesting that 
superconductivity is associated with the quantum fluctuations.
For $t^\prime=0.1$ and $t^\prime=0.3$, the SC dome
contains the point where $T^*\to 0$, but is {\em{not}} centered around $n_c$.
Instead, the second-order point is found on the low-doping 
side of the SC dome.  Note that 
the maximum SC transition temperature increases slightly
with increasing $t^\prime$ which is in agreement with previous results for 
a four-site cluster, and the same interaction strength~\cite{e_khatami_08}.

\section*{Discussion}

A detailed study of the phase diagram of the 2D Hubbard model with
next-near-neighbor hopping has allowed us to  identify the nature of 
the QCP under the SC dome.
We argue that QCP is the terminal point of a line of second-order critical 
points associated with  first-order phase-separation transitions. The critical
temperature is driven to zero  as $t^\prime \to 0$.
For positive $t'$, a Mott liquid and a Mott gas coexist at fixed $\mu$~\cite{alex_PS}.
In real materials other parameters, such as electron-phonon interaction, inter-site 
electron-electron interaction or inhomogeneities, might play a role similar to $t'$. 

Generally, it is accepted that the model with $t'>0$ describes
the electron-doped and that with $t'<0$ described the hole-doped
cuprates. We find that the model for $t'>0$ does not display
quantum criticality, but rather classical criticality.
The QCP is found only for $t'=0$ and as is known from
other quantum critical systems, it will strongly affect the
system for a wide range of parameters and temperatures
around this point, including $t'<0$, the model for the hole-doped
cuprates.

The relationship of the superconductivity with the QCP at
$t^\prime=0$ is not yet clear, but the fact that the dome is centered at
the QCP suggests that the incipient phase separation creates
conditions favorable for superconductivity~\cite{Kivelson_PS}.  The
Mott liquid phase may provide regions where the spin-mediated pairing
interaction is strong and the Mott gas may provide 
regions where there are quasiparticles to pair. It may also be that 
the incipient charge fluctuations when combined with the 
antiferromagnetic spin fluctuations enhance the pairing 
within a narrow region near the QCP~\cite{Rome}.  
The phase-separated region might also be related to the pervasive
inhomogeneities observed in cuprates which
led to theoretical scenarios for an 
inhomogeneity-based pairing mechanism~\cite{e_arrigoni_04,e_carlson_04} 
or an enhancement of pairing interactions~\cite{i_martin_05,k_aryanpour_06a,w_feng_06,y_loh_2007,t_maier_09}.
Finally, it has been
suggested that in the vicinity of the QCP, even a weak retarded 
attractive interaction may become far more effective at inducing 
pairing~\cite{she}. The relation between the QCP and superconductivity
will be explored in greater detail in future studies.

\section*{Conclusion}

We find that the QCP at $t^\prime=0$  
of the 2D Hubbard model 
is the zero-temperature limit
of a line of second-order critical points associated with a first-order 
phase separation transition, which occur at finite temperature 
when $t^\prime>0$. The filling associated with the second-order 
critical point is determined
from the peak position in the compressibility versus chemical potential. The peak 
grows by decreasing the temperature or increasing the cluster size or the interaction strength. We 
also show that for $t^\prime>0$ and at $n=n_c$,  
the charge susceptibility diverges at a finite 
temperature which decreases by decreasing $t^\prime$. As $t^\prime\to 0$, $n_c$ moves continuously
from values close to half filling to the filling that corresponds to the QCP.
For $t^\prime=0$, the SC dome is centered at the QCP where the pseudogap 
temperature, $T^*$, also 
vanishes. This suggests that the incipient phase separation might play a role in the pairing 
mechanism. However, for $t^\prime>0$, while $T^*$ seems to vanish roughly at the 
center of the dome, phase separation happens at temperatures much higher than the SC 
temperature with classical critical points that move to the lower-doping side 
of the dome.

\section*{Acknowledgments} 
We would like to thank P.\ Phillips, S.\ Kivelson, C.\ Varma,
and J.\ Zaanen for useful conversations.  This research was supported by 
NSF under Grant No. DMR-0706379.  J.M. and M.J. are supported by the NSF PIRE under
Project No. OISE-0730290.
M.J. and R.T.S. are also supported by DOE SciDAC under Project No. DE-FC02-06ER25792.
This research used resources of the National Center for
Computational Sciences at Oak Ridge National Laboratory, which is supported 
by the Office of Science of the U.S. Department of Energy under Contract 
No. DE-AC05-00OR22725.

\end{document}